\documentclass[fleqn,10pt]{wlscirep}
\usepackage{amsmath,amssymb,graphicx}
\usepackage{epstopdf}
\usepackage{bm}
\newcommand{\triI}[0]{\int\hspace{-2.5mm}\int\hspace{-2.5mm}\int\!\!}%%closely spaced triple integral
\newcommand{\ab}[1]{\mathrm{#1}}
\title{Super-resolution imaging using the spatial-frequency filtered intensity fluctuation correlation}
\author[1,*]{Jane Sprigg}
\author{Tao Peng}
\author{Yanhua Shih}
\affil{University of Maryland, Baltimore County - 1000 Hilltop Circle, Baltimore MD, 21250, United States of America}
\affil[*]{jane.sprigg@gmail.com}
\begin{abstract}
We report an experimental demonstration of a nonclassical imaging mechanism with super-resolving power beyond the Rayleigh limit. When the classical image is completely blurred out due to the use of a small imaging lens, by taking advantage of the intensity fluctuation correlation of thermal light, the demonstrated camera recovered the image of the resolution testing gauge. This method could be adapted to long distance imaging, such as satellite imaging, which requires large diameter camera lenses to achieve high image resolution.
\end{abstract}
\begin{document}
\flushbottom
\maketitle
\thispagestyle{empty}
\section*{Introduction}
Improving the resolution of optical imaging has been a popular research topic in recent years\cite{BrambillaLugiato2005,MacconeShapiro2009,TisaZappa2010,GongHan2012, GongHan2013,Oh:13}. A commonly used simple approach is to measure the autocorrelation of two identical classical images, effectively squaring the classical image, $\langle I_1(\bm{\rho}_1) \rangle \langle I_1(\bm{\rho}_1) \rangle$, where $\bm{\rho}_1$ is the transverse coordinate of the detector. This autocorrelation produces a maximum $\sqrt{2}$ gain of the spatial resolution. However, the imaging resolution of such a setup can be further improved by changing the measurement from $\langle I_1(\bm{\rho}_1) \rangle \langle I_1(\bm{\rho}_1) \rangle$, in terms of intensity, or $\langle n_1(\bm{\rho}_1) \rangle \langle n_1(\bm{\rho}_1) \rangle$, in terms of photon number counting, to the intensity fluctuation correlation $\langle \Delta I_1(\bm{\rho}_{1}) \Delta I_2(\bm{\rho}_{2}) \rangle$, or $\langle \Delta n_1(\bm{\rho}_{1}) \Delta n_2(\bm{\rho}_{2}) \rangle$, where $\bm{\rho}_1$ and $\bm{\rho}_2$ are the transverse coordinates of two spatially separated detectors. Then, if only those fluctuation correlations due to the higher spatial frequencies from $\Delta I_2(\bm{\rho}_{2})$ are selected, a super-resolving image can be observed from the joint detection of the intensity fluctuations at the two detectors. The physics behind this super-resolution is similar to the original thermal light ghost imaging \cite{thermalGI,ScarcelliBerardi2006}, and is quite different from an autocorrelation measurement. It should be emphasized that the reported result is also different than that of Oh et. al.\cite{Oh:13}; while the authors measured the intensity fluctuations, it was the intensity fluctuation autocorrelation $\langle \Delta I_1(\bm{\rho}_{1})^2 \rangle$, which was still limited by the $\sqrt{2}$ resolution improvement of an autocorrelation measurement.

In this Report, we demonstrate a camera with resolution beyond the classical Rayleigh limit. Similar to the original thermal ghost imaging experiments\cite{thermalGI,ScarcelliBerardi2006}, the camera produces an image by the measurement of $\langle \Delta I_1(\bm{\rho}_{1}) \Delta I_2(\bm{\rho}_{2}) \rangle$; the camera consists of a typical imaging setup, except it has two sets of independent and spatially separated detectors: $D_1$ placed on the image plane, and $D_2$ placed on the Fourier transform plane. Crucially, $D_2$ integrates (sums) only the higher spatial frequencies, or transverse wavevectors, by blocking the central area of the Fourier transform plane. The image is calculated from the intensity fluctuations of $D_1$, at each transverse position $\bm{\rho}_{1}$, and the bucket detector $D_2$. The measurement can be formulated as $\Delta R_c(\bm{\rho}_1)= \langle \Delta I_1(\bm{\rho}_1) \int d \bm{\kappa}_2 \ab{F}(\bm{\kappa}_2) \Delta I_2(\bm{\kappa}_2) \rangle$, where $\ab{F}(\bm{\kappa}_2)$ is a filter function which selects the higher spatial frequencies. 
\section*{Experimental setup}
\begin{figure}
\center
		\includegraphics[width=4in]{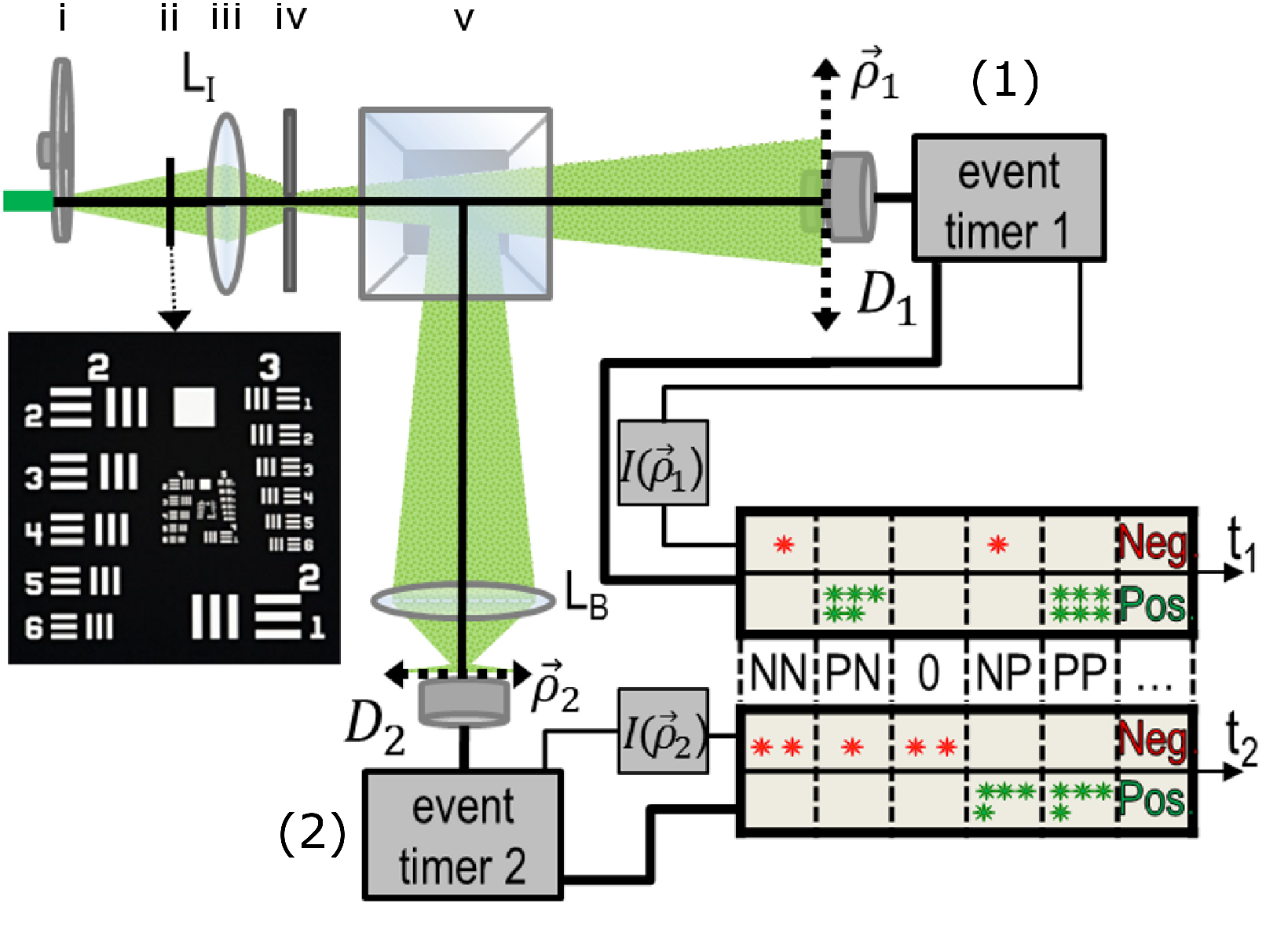}
	\caption{Experimental setup: a 10 mm diameter 532 nm wavelength laser beam scatters from a rotating ground glass (i) and strikes a 1951 USAF Resolution Testing Gauge (ii), then the imaging lens $L_I$ (iii) and a pinhole of $\approx$1.36 mm diameter (iv). The light splits to arm one, where it is collected by a scanning point detector $D_1$, and arm two, where it is collected by a spatially filtered bucket detector, consisting of a lens $L_B$ located on the image plane and a multimode fiber tip in the Fourier transform plane. The image is then calculated using the Photon Number Fluctuation Correlation (PNFC) circuit.
}\label{fig.1}
\center
\end{figure}
\begin{figure}
\center
		\includegraphics[width=\columnwidth]{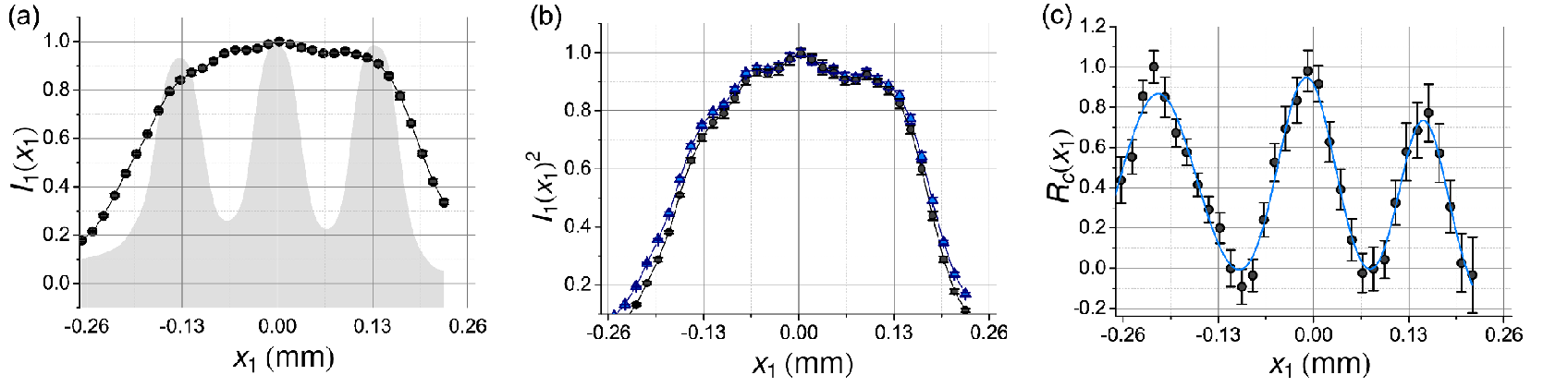}
	\caption{Resolution comparison for different imaging methods of three 0.01241 mm wide slits imaged by a 10 mm diameter source: (a) unresolved first-order classical image, where the gray shading marks the location of the slits; (b) unresolved images from the fluctuation autocorrelation. The black dots indicates $[I_1(x_1)]^2$ and the blue triangles show $[\Delta I_1(x_1)]^2$, as seen in Oh et. al.\cite{Oh:13}; (c) completely resolved image observed from $\Delta R_c(x_1)$ where the solid line is a Gaussian data fitting.}
	\label{fig.2} 
	\center
\end{figure}

Fig. \ref{fig.1} illustrates the laboratory-demonstrated camera setup. We used a standard narrow spectral bandwidth pseudo-thermal light source consisting of a 10 mm diameter 532 nm wavelength laser beam scattered by millions of tiny diffusers on the surface of a rotating ground glass. The object imaged was the 5-3 element of a 1951 USAF Resolution Testing Gauge. Like a traditional camera, the imaging lens, $L_I$, had an aperture limited by an adjustable pinhole to approximately 1.36 mm diameter. However, this imaging device has two optical arms behind its imaging lens $L_I$. The light transmitted by the object falls on a single-mode 0.005 mm diameter fiber tip that scans the image plane of arm one and is then interfaced with a photon counting detector $D_1$. The combination of the scanning fiber tip, $D_1$, and photon counting detector acts as a CCD array. In arm two, the second lens $L_B$ is placed behind the image plane and performs a Fourier transform of the field distribution of the image plane of arm two. $D_2$, a scannable multimode 0.105 mm diameter fiber interfaced with a photon counting detector, is placed in the Fourier transform plane of $L_B$ and integrates only the higher spatial frequencies while filtering out the lower spatial frequency modes; we emphasize that the placement of this ``spatial filter'' does not depend on knowledge of the Fourier transform of the image.

The intensity fluctuations in this experiment were recorded by a Photon Number Fluctuation Correlation (PNFC) circuit\cite{pnfChen,huiThesis}, which independently records the arrival time of each photo-detection event at $D_1$ or $D_2$. The intensity, measured by the number of photons detected per second, is divided into a sequence of short time windows, each of which needs to be less than the second-order coherence time of the light; it is important that the width of the time window not be too long. The software first calculates the average 
intensity per short time window, $\bar{n}_{s}$, where s=1,2 indicates the detector, and then the difference or fluctuation term for each time window: $\Delta n_{j,s}=n_{j,s}-\bar{n}_s$. The corresponding statistical average of $\langle \Delta n_{1} \Delta n_{2} \rangle$ is thus
\begin{align}\label{NormCorrelations}
&\langle \Delta n_1 \Delta n_2 \rangle= \frac{1}{N} \Big{[} \sum_{j} (\Delta n_{j,1} \Delta n_{j,2})\Big{]}.
\end{align} 
It should be emphasized that when the fluctuation correlation is calculated between the two detectors, $\Delta n_1$ and $\Delta n_2$ have not yet been time-averaged. The time averaging is performed after the correlation, as indicated by the multiplication appearing inside the sum over $j$. 
\section*{Experimental results}

Typical experimental results are presented in Fig. \ref{fig.2}. In this measurement, the 5-3 element of a 1951 USAF Resolution Test Target was imaged in one dimension by scanning $D_1$ in the x-direction along the slits. Fig. \ref{fig.2}(a) shows a completely unresolved classical image of the three slits, $I_1(x_1)$, that was directly measured by the scanning detector $D_1$. For reference the gray shading indicates the location of the slits. Fig. \ref{fig.2}(b) shows two results: the black dots plot the autocorrelation $[I_1(x_1)]^2$, while the blue triangles show the fluctuation autocorrelation $\langle [\Delta I_1(x_1)]^2\rangle$ at each point. $\langle [\Delta I_1(x_1)]^2\rangle$ was calculated from the intensity fluctuation autocorrelation of $D_1$. The measurements in Fig. \ref{fig.2}(b) have a $\sqrt{2}$ resolution gain, similar to that of Oh et. al. \cite{Oh:13}. Using the Rayleigh limit \cite{Hecht,goodmanF} $\delta x=1.22\lambda\,s_I /D$,\cite{Hecht,goodmanF} where $s_I$ is the distance from lens $L_I$ to the image plane and $\lambda$ is the wavelength of illumination, 
the expected resolution of the autocorrelation in the image plane is approximately $\delta x/\sqrt{2}=0.13$ mm which, as seen in \ref{fig.2}(b), is not enough to resolve the three slits which have a slit-to-slit separation of about 0.13 mm. However, by spatially filtering arm two, the three slits of the 5-3 element of the gauge were clearly separated when correlated with arm 1, as seen in Fig. \ref{fig.2}(c). The error bars in Figs. \ref{fig.2}(a) and (b) are quite small, especially when compared to those in Fig. \ref{fig.2}(c). This is a typical negative feature of second-order measurements; compared to first-order classical imaging, in order to achieve the same level of statistics the reported imaging mechanism needs a longer exposure time. How much longer the measurement takes depends on the power of the light source and other experimental parameters.

Fig. \ref{fig.2}(c) is the sum of two measurements obtained by placing the bucket fiber tip at two points in the Fourier transform plane: $x_{2+}=0.22$ mm and $x_{2-}=-0.24$ mm; this selects the higher spatial frequency modes which fall onto the two fiber tips and ``blocks'' all other spatial frequency modes. We represent this mathematically with the filter function $F(x_2)=\mathrm{\Pi}(x_2-x_{2+},D_F)+\mathrm{\Pi}(x_2-x_{2-},D_F)$ to simulate the physical ``spatial filtering'', where $\mathrm{\Pi}(x,w)$ is a rectangle function of width w, $D_F=0.105$ mm is the fiber diameter, and $x_2$ is measured from the central maximum. Then in one dimension for $\bm{\kappa}\propto k x_2/f_B$, where $f_B$ is the focal length of the bucket lens, $\Delta R_c(x_1)=\langle \Delta I_1(x_1) \int \ab{d}x_2 \ab{F}(x_2)\Delta I_2(x_2) \rangle$, and only integrates the higher spatial frequencies collected by the bucket detector in the neighborhood of $x_2=x_{2+},x_{2-}$. Again, this ``spatial filtering'' does not require any knowledge of the object or its Fourier transform function. 

One way to improve these results is to replace $D_2$ with a CCD array; the CCD would still be in the Fourier transform plane, but with the central pixels blocked. This would allow $D_2$ to collect more light of higher spatial frequencies. Although the limits of our equipment, software data storage, and time constraints prevented the authors from making such improvements, in this reported measurement all three slits of the resolution gauge are certainly well-resolved, while both the classical imaging and the autocorrelation mechanisms could not resolve it. 
%demonstrating a resolution improvement greater than that of an autocorrelation mechanism. 

\section*{Discussion and theory} 

In the experiment, we use the spatial correlation of the noise, $\langle \Delta I_1(\mathbf{r}_1, t_1) \Delta I_2(\mathbf{r}_2, t_2) \rangle$, to produce an image from the joint photo-detection of two independent and spatially separated photodetectors, $D_1$ and $D_2$. In the following, we outline the theory behind our experiment. First we briefly consider how a first-order or classical camera produces an image in its image plane, $\langle I_1(\bm{\rho}_{1}) \rangle$. 

The experiment was performed using a pseudothermal light source, created by placing a rotating ground glass in the path of a laser beam. The ground glass contains a large number of tiny scattering diffusers which act as sub-sources. The wavepackets of scattered light play the role of subfields; each diffuser scatters a subfield to all possible directions, resulting in the subfields acquiring random phases. Each sub-field propagates from the source plane to the image plane by means of a propagator or Green's function, $E_m(\bm{\rho}_{1}) = E_m \, g_m(\bm{\rho}_{1})$, where $E_m$ is the initial phase and amplitude of the field emitted by sub-source $m$ and $g_m(\bm{\rho}_{1})$ is the Green's function which propagates the light from the $m^{th}$ sub-source located at $\bm{\rho}_m$ to the point $\bm{\rho}_1$ at some distance $z$ from the source plane. To simplify the problem, we assume the fields are monochromatic and ignore the temporal part of the propagator. Then the light measured at coordinate $(\mathbf{r}, t)$ is the result of the superposition of a large number of subfields, $\sum^\infty_{m=1}E_m(\mathbf{r}, t)$, each emitted from a point sub-source, 
\begin{align}\label{Image-00}
I(\mathbf{r}, t) &= \sum_{m} E_m^*(\mathbf{r}, t) \sum_n E_{n}(\mathbf{r}, t) 
= \sum_{m} \Big{|} E_m(\mathbf{r}, t) \Big{|}^2 
+ \sum_{m\neq n} E_m^*(\mathbf{r}, t) E_{n}(\mathbf{r}, t) 
= \langle I(\mathbf{r}, t) \rangle + \Delta I(\mathbf{r}, t) 
\end{align}
where $\langle I(\mathbf{r}, t) \rangle$, the mean intensity, is the result of the $m$th subfield interfering with itself; $\Delta I(\mathbf{r}, t)$, the intensity fluctuation, is the result of the $m$th subfield interfering with the $n$th subfield, $m \neq n$, and is usually considered noise because $\langle \Delta I(\mathbf{r}, t) \rangle =0$ when taking into account all possible random phases of the subfields. 

A classical imaging system measures the mean intensity distribution on the image plane, $\langle I_1(\bm{\rho}_1) \rangle$, where we have assumed a point detector $D_1$ is placed at coordinate $\bm{\rho}_1$, the transverse coordinate of the image plane. In an ideal imaging system, the self-interference of subfields produces a perfect point-to-point image-forming function. The ideal classical image assuming an infinite lens is the convolution between the aperture function of the object $|A(\bm{\rho}_{O})|^2$ and the image-forming $\delta$-function which characterizes the point-to-point relationship between the object plane and the image plane\cite{Hecht,goodmanF,shih}.
\begin{align}\label{Image-Eq}
\langle I(\bm{\rho}_1) \rangle &= \sum_{m} \Big{|} E_m \int d \bm{\rho}_{O} \, g_m(\bm{\rho}_{O})  A({\bm{\rho}_{O}}) g_O(\bm{\rho}_{1}) \Big{|}^2 
\propto \int_{obj} d\bm{\rho}_{O} \, |A(\bm{\rho}_{O})|^2 \, \delta(\bm{\rho}_{O} + \frac{\bm{\rho}_{1}}{\mu})
= |A(-\bm{\rho}_{1} / \mu)|^2
\end{align}
where $\mu = s_I / s_O$ is the magnification factor, $g_m(\bm{\rho}_O)$ is a Green's function propagating the $m^{th}$ subfield from the source plane to the object plane over a distance $z$ , and $g_O(\bm{\rho}_1)$ is a function propagating the subfield from the object plane to the detection plane over a distance $s_O+s_I$, and including the imaging lens. $A({\bm{\rho}_{O}})$ is an arbitrary function describing the object aperture.

In reality, due to the finite size of the imaging system, we rarely have a perfect point-to-point correspondence.   
Incomplete constructive-destructive interference blurs the point-to-point correspondence to point-to-spot
correspondence.  The $\delta$-function in the convolution of Eq.~\ref{Image-Eq} is then replaced by a point-to-spot image-forming function, or a point-spread function which is determined by the shape and size of the lens. For a lens with a finite diameter, one common model describes the shape or pupil of the lens as a disk of diameter D:
\begin{equation}\label{Image-Eq-2}
\langle I_1(\bm{\rho}_1) \rangle =
\int_{obj} d\bm{\rho}_{O} \, |A(\bm{\rho}_{O})|^2 \,
\textrm{somb}^2\big{[} \frac{\pi}{\lambda} \frac{D}{s_O}\,\big{|}\bm{\rho}_{O} + 
\frac{\bm{\rho}_{1}}{\mu} \big{|} \big{]} 
\end{equation}
where the sombrero-like point-spread function is defined as $\textrm{somb}(x)\equiv 2 J_1(x) /x$; $J_1(x)$ is the first-order Bessel function. The image resolution is determined by the width of the somb-function: the narrower the higher. A larger diameter lens results in a narrower somb-function and thus produces images with higher spatial resolution. 

To simplify the mathematics, it is common to approximate a finite lens as a Gaussian $\ab{e}^{-(\rho_L/(D/2))^2}$ with diameter $D$, but a smoother falloff than the disk approximation. This leads to a Gaussian imaging-forming function:
\begin{equation}\label{Image-Eq-2G}
\langle I_1(\bm{\rho}_1) \rangle =
\int_{obj} d\bm{\rho}_{O} \, |A(\bm{\rho}_{O})|^2 \,
\textrm{exp}\big{[} -(\frac{\pi}{\lambda} \frac{D}{2 s_O}\,\big{|}\bm{\rho}_{O} + 
\frac{\bm{\rho}_{1}}{\mu} \big{|})^2 \big{]}. 
\end{equation}
This Gaussian version of the imaging equation will be used later in numerical calculations to simplify the mathematical evaluation. 

It is clear from Eqs. \ref{Image-Eq-2} and \ref{Image-Eq-2G} that for a chosen value of distance $s_O$, a larger imaging lens and shorter wavelength will result in a narrower point-spread function, and thus a higher spatial resolution of the image.  

Now we consider the noise produced image that is observed from the measurement of Fig. \ref{fig.1} by means of $\langle \Delta I_1(\bm{\rho}_1)\Delta I_2 (\bm{\rho}_2)\rangle$.  To make the explanation of the experimental results easier to follow, first we examine the case where two point scanning detectors $D_1$ and $D_2$ are placed in the image planes of arm one and arm two:
\begin{align}
\langle \Delta I_1(\bm{\rho}_{1}) \Delta I_2(\bm{\rho}_{2}) \rangle&=\big\langle \sum_{m \neq n} E_m^*(\bm{\rho}_{1}) E_n(\bm{\rho}_{1}) \sum_{p \neq q}  E^*_p(\bm{\rho}_{2}) E_q(\bm{\rho}_{2}) \big\rangle 
= \sum_{m = q} E_m^*(\bm{\rho}_{1}) E_m(\bm{\rho}_{2}) \sum_{n = p} E_n(\bm{\rho}_{1}) E^*_n(\bm{\rho}_{2}) \cr
&\simeq \Big{|} \sum_{m} E_m^*(\bm{\rho}_{1}) E_m(\bm{\rho}_{2}) \Big{|}^2
\end{align}
The calculation of $\sum_{m} E_m^*(\bm{\rho}_{1}) E_m(\bm{\rho}_{2})$ is straightforward: 
\begin{align}
\sum_{m} E_m^*(\bm{\rho}_{1}) E_m(\bm{\rho}_{2})&= \sum_{m} \Big{[} E_m^* \int d \bm{\rho}_{O} g^*_m(\bm{\rho}_{O}) \int d \bm{\kappa} A^*(\bm{\kappa}, {\bm{\rho}_{O}}) g^*_O(\bm{\kappa}, \bm{\rho}_{1}) \Big{]}\Big{[} E_m \int d \bm{\rho}_{O'}  g_m(\bm{\rho}_{O'}) \int d \bm{\kappa}' A(\bm{\kappa}', {\bm{\rho}_{O'}}) g_{O'}(\bm{\kappa}', \bm{\rho}_{2}) \Big{]} \cr
&= \sum_{m} E_m^* \int d \bm{\rho}_{O} \int d \bm{\rho}_{O'} g^*_m(\bm{\rho}_{O}) E_m g_m(\bm{\rho}_{O'}) \Big{[} \int d \bm{\kappa}  \, A^*(\bm{\kappa}, {\bm{\rho}_{O}}) \, e^{-i \bm{\kappa} \cdot \bm{\rho}_{O}}\textrm{somb} \big{[}\frac{\pi}{\lambda} \frac{D}{s_O} \big{|}\bm{\rho}_{O} + 
\frac{\bm{\rho}_{1}}{\mu} \big{|} \big{]} \Big{]} \cr
& \times \Big{[} \int d \bm{\kappa}'  A(\bm{\kappa}', {\bm{\rho}_{O'}}) e^{i \bm{\kappa}' \cdot \bm{\rho}_{O'}}\textrm{somb} \big{[} \frac{\pi}{\lambda} \frac{D}{s_O} \big{|}\bm{\rho}_{O'} + 
\frac{\bm{\rho}_{2}}{\mu} \big{|} \big{]} e^{-i k(z_0+s_O)/(2z_0 s_O)(\rho_{O}^2-\rho_{O'}^2)}e^{-i k/(2s_I)(\rho_{1}^2-\rho_{2}^2)} \Big{]}
\end{align}

Next, we complete the summation over $m$ in terms of the subfields, or the sub-sources, by means of an integral over the entire source plane. This integral results in the well-known Hanbury-Brown Twiss (HBT) correlation: $\textrm{somb}^2[(\pi \Delta\theta)/\lambda| \bm{\rho}_{O} - \bm{\rho}_{O'}|]$, where $\Delta \theta$ is the angular diameter of the light source relative to the object plane. To simplify further calculations, we assume a large value of $\Delta \theta$ and approximate the somb-function to a $\delta$-function evaluated at $\bm{\rho}_{O} = \bm{\rho}_{O'}$, $\bm{\kappa} = \bm{\kappa}'$. $\langle \Delta I_1(\bm{\rho}_{1}) \Delta I_2(\bm{\rho}_{2}) \rangle$ is therefore approximately equal to: 
\begin{align}
\label{sqImg}
\langle \Delta I_1(\bm{\rho}_{1}) \Delta I_2(\bm{\rho}_{2}) \rangle&\approx\Big{|} \int d \bm{\rho}_{O} |A({\bm{\rho}_{O}})|^2 \textrm{somb} \big{[}\frac{\pi}{\lambda} \frac{D}{s_O} \, \big{|}\bm{\rho}_{O} + 
\frac{\bm{\rho}_{1}}{\mu} \big{|} \big{]}\textrm{somb} \big{[} \frac{\pi}{\lambda} \frac{D}{s_O} \big{|}\bm{\rho}_{O} + 
\frac{\bm{\rho}_{2}}{\mu} \big{|}\big{]} e^{-i k/(2s_I)(\rho_{1}^2-\rho_{2}^2)}\Big{|}^2
\end{align}
It is clear that when $\bm{\rho}_1=\bm{\rho}_2$ in Eq. \ref{sqImg}, the measurement of $\langle \Delta I_1(\bm{\rho}_1)\Delta I_2(\bm{\rho}_1)\rangle$ produces an image with a $\sqrt{2}$ resolution gain, with an imaging resolution due to the image-forming $\textrm{somb}$-functions, i.e., $\bm{\rho}_1=\bm{\rho}_2\simeq\mu\bm{\rho}_O$. When the lens is large enough to resolve the object, the result is a point-to-point reproduction of the image only when $\bm{\rho}_1=\bm{\rho}_2$; otherwise for small lens apertures Eq. \ref{sqImg} forms a point-to-spot image when $|\bm{\rho}_O+\bm{\rho_1}/\mu|<\lambda s_O/D$ and $|\bm{\rho}_O+\bm{\rho_2}/\mu|<\lambda s_O/D$. 

Now we move $D_2$ to the Fourier transform plane of $L_B$ of arm two, i.e., to its focal plane, effectively performing a Fourier transform of the field distribution of the image plane. In addition, $D_2$ is placed off-center relative to the optic axis of the lens to select part of the spatial frequencies on the Fourier transform plane, acting as a spatial frequency filter. Mathematically, 

\begin{align}\label{filter}
\Delta R_c(\bm{\rho}_{1})&= \Big{\langle} \Delta I_1(\bm{\rho}_{1}) \int d \bm{\rho}_{2} \Delta I_2(\bm{\rho}_{2}) \Big{\rangle} \cr
&\propto \Big{|} \left(\triI d \bm{\kappa}_{2}\ab{F}(\bm{\kappa}_2)d \bm{\rho}_{2}d \bm{\rho}_{O} \, |\ab{A}({\bm{\rho}_{O}})|^2 \,\textrm{somb} [\frac{\pi}{\lambda} \frac{D}{s_O} \,\big{|}\bm{\rho}_{O} + 
\frac{\bm{\rho}_{1}}{\mu}\big{|}]\right.\left.\textrm{somb} [ \frac{\pi}{\lambda} \frac{D}{s_O} \, \big{|}\bm{\rho}_{O} + 
\frac{\bm{\rho}_{2}}{\mu} \big{|} ]\ab{e}^{-\ab{i} k/(2s_I)(\rho_{1}^2-\rho_{2}^2)}\right.\cr
&\times\left.\ab{e}^{-\ab{i}\bm{\kappa}_2 \cdot \bm{\rho}_2}\right) \Big{|}^2.  
\end{align}
As a result of the spatial filter function $\ab{F}(\bm{\kappa}_2)$, the imaging resolution of Eq. \ref{filter} is much narrower than that of a first-order image; however, it is difficult to simplify this equation further in this form.

To get a better understanding of the physics behind Eq. \ref{filter}, instead of modeling the finite radius of the lens as a disk, which results in the somb-function, we approximate the finite radius of the lens as the Gaussian function $\ab{e}^{-x_L^2/(D/2)^2}$ with a half-width $D/2$, and evaluate in one dimension. This leads to a Gaussian imaging-forming function instead of the somb-function.  Working in one dimension, we change $\bm{\rho}_O$ to $x_O$; $\bm{\rho}_1$ to $x_1$; $\bm{\rho_2}$ to $x_2$, etc. Then $\Delta R_c(x_{1})$ simplifies to:
\begin{align}
\label{filter1d}
\Delta R_c(x_{1})&\approx\left|\int_{-\infty}^{\infty}\!\!\ab{d}x_O|\ab{A}(x_O)|^2 \ab{e}^{-\ab{i}\frac{k}{2s_{\!I}}x_1^2}\ab{e}^{-(\frac{k D}{4s_O})^2(x_O+x_1/\mu)^2}\!\int_{-\infty}^{\infty}\hspace{-2.5mm}\int_{-\infty}^{\infty}\! \ab{d}\kappa_2\ab{d}x_2\ab{F}(\kappa_2)\ab{e}^{-(\frac{k D}{4s_O})^2(x_O+x_2/\mu)^2}\ab{e}^{\ab{i}\frac{k}{2s_{\!I}}x_2^2}\ab{e}^{\ab{-i}\kappa_2 x_2}\right|^2.
\end{align}
Corresponding to the experimental measurement, where $D_2$ was placed at two off-center points in the Fourier transform plane, we model the filter function in one dimension by two rectangle functions: $F(\kappa_2)=\Pi(x_F-x_{2+},D_F)+\Pi(x_F-x_{2-},D_F)$, where $\kappa_2=\frac{k x_F}{f_B}$, while $\ab{A}(x_O)=\Pi(x_O,w)+\Pi(x_O-2w,w)+\Pi(x_O+2w,w)$; assuming $x_2$, $x_O$, and $\kappa_2(x_F)$ are integrated from $\pm\infty$, the resulting equation is an analytic expression. Define
\begin{align}
f(x_1,x_O)=\ab{exp}\left[-\left(\frac{k D}{4 s_O}\right)^2\left(\frac{x_1^2}{\mu^2}+2 \frac{x_1 x_O}{\mu}+\frac{8 \mu^2+(k D^2 x_O^2/(4 s_O))^2}{4\mu^2+(k D^2/(4 s_O))^2}\right)\right] \ab{exp}\left[\ab{i}\left(\frac{k x_1^2}{2 \mu s_O}-\frac{k^3 D^4 x_O^2}{2 k^2 D^4+128 \mu^2 s_O^3}\right)\right]|\ab{A}(x_O)|^2 ,
\end{align}
which contains the imaging equation in Gaussian form for $x_1$ and $x_2$, in addition to some phase terms which are not observable in a first-order image, and
\begin{align}
a=\frac{-\mu^2 s_O^2}{k(k(D/2)^2+2\ab{i}\mu s_O)},\hspace{3mm} b(x_O)=\frac{\ab{i}k \mu D^2 x_O}{k D^2+8\ab{i}\mu s_O};
\end{align}
Then Eq. \ref{filter1d} is, after evaluating the $\kappa_2$ and $x_2$ integrals,
\begin{align}\label{expSym}
\Delta R_c(x_{1})&=\left|\int\!\ab{d}x_O f(x_1,x_O)\ab{e}^{-\frac{b(x_O)^2}{4 a}}\Big(\big(\ab{Erfi}(\frac{b(x_O)+2a k(\frac{D_F}{2 f_B}-\frac{x_{2-}}{f_B})}{2\sqrt{a}})-\ab{Erfi}(\frac{b(x_O)-2a k(\frac{D_F}{2 f_B}+\frac{x_{2-}}{f_B})}{2\sqrt{a}})\big)\right.\cr
&\left.+\big(\ab{Erfi}(\frac{b(x_O)+2a k(\frac{D_F}{2 f_B}-\frac{x_{2+}}{f_B})}{2\sqrt{a}})-\ab{Erfi}(\frac{b(x_O)-2a k(\frac{D_F}{2 f_B}+\frac{x_{2+}}{f_B})}{2\sqrt{a}})\big)\Big)\right|^2.
\end{align}
Then it is easy to see that restricting the allowed spatial frequencies of the Erfi functions constrains the values $x_O$ is allowed to take, which, together with $f(x_1,x_O)$, improves the ability to resolve different points on the object plane. However, without evaluation Eq. \ref{expSym} may still not be clear enough to show exactly how the resolution is affected, so we have included the following figures which plot some informative values to support our experimental observation.
\begin{figure}
		\includegraphics[width=\columnwidth]{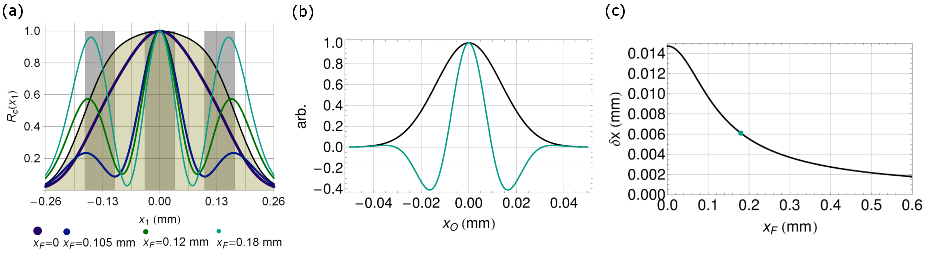}
	\caption{All plots were calculated from Eq. \ref{expSym}. (a)\, \,Theoretical comparison of the resolution of a 1.36 mm lens for three slits with a width and separation of 0.01241 mm. The transparent gold shading shows the first-order unresolved image, while the light gray indicates the ideal image; the lines plot Eq. \ref{expSym} with spatial filter $\ab{F}(\kappa_2)$ for several cutoff values $\kappa_2=\frac{k x_2}{f_B}$, $x_{2\pm}=\pm x_F$, where k is the wavenumber and $f_B$ is the focal length of lens $L_B$, versus the position of $D_1$ in the image plane. As the cutoff frequency increases the ability to resolve the three slits also increases. \,\, (b) Comparison of the first-order Gaussian imaging function (black) to the second-order imaging function (teal) for $x_F$=0.18mm. The y-axis is in arbitrary units of normalized intensity. \,\,(c) Theoretical comparison of the resolution of Eq. \ref{expSym} for a 1.36 mm lens as the cutoff frequency $x_F$ increases. The dot marks the second-order imaging resolution of $x_F=0.18$ mm from (a); the first-order imaging resolution corresponds to $x_F=0$ }
	\label{fig.3}
\end{figure}
%%%
Fig. \ref{fig.3}(a) compares the theoretical first-order unresolved image of three slits with the second-order fluctuation correlation image calculated using Eq. \ref{expSym}. The filter function is calculated for a fiber diameter $D_F$ of 0.105 mm and varying distances $x_F$ from the center of the Fourier transform plane. Note that the plot for $x_F=0.18$ mm demonstrates similar behavior to the observed experimental data, including the shift of the left and right peaks away from center. It is clear that, for a lens diameter of 1.36 mm, the gold transparent plot is completely unresolved. However, as the cutoff frequency $\kappa_2=k x_F/f_B$ increases, the second-order resolution also increases as seen in the increasing separation of the peaks in Fig. \ref{fig.3}. This is more clearly illustrated in Figs. \ref{fig.3}(b) and (c); in (b) the imaging function of the first-order image (black) is plotted with the second-order imaging function (teal) for $x_F=0.18$mm. It is clear that the second-order imaging function is much narrower. In (c) the half-width of the imaging function in Eq. \ref{expSym} is calculated at $x_1$=0 for increasing values of $x_F$. Using the estimated experimental parameters, the second-order imaging resolution starts equal to the first order at $x_F=0$ and increases to the experimental setup's limit of about 0.002 mm. 

It is evident from the experiment and theoretical calculations that the increase in spatial resolution is strongly dependent on the chosen spatial filter. It is, in effect, applying a high-pass spatial filter to one arm, producing an edge-sharpening effect \cite{goodmanF,Hecht}. The interesting part is that the correlation of the spatially filtered intensity fluctuations with arm one produces a resolved image, especially since neither arm ``sees'' a resolved image. This correlation filters out the lower spatial frequencies of the unresolved image of arm one, yielding a resolved image in the intensity fluctuation correlation rather than the intensity.
\section*{Conclusion}

In summary, by using a high-pass spatial filter in the non-resolving side of a two-arm camera, the measurement of the intensity fluctuation correlation $\langle\Delta I_1(\bm{\rho}_1) \Delta I_2(\bm{\rho}_2) \rangle$ was able to resolve an object that could not be resolved by a traditional camera. This imaging method would be particularly useful for long-distance imaging in situations where it is impractical to have large lenses but high resolution is still desired, as it could take advantage of the large angular size of the sun, $0.5^{\circ}$ relative to the earth, and the correspondingly small coherence length, on the order of 0.2 mm. In addition, since the thermal light image in $\langle\Delta I_1(\bm{\rho}_1) \Delta I_2(\bm{\rho}_2) \rangle$ is in general turbulence-free\cite{meyersturb}, this method would be particularly attractive for satellite cameras taking high resolution images of objects on the ground.  Technically more complicated optics or electronics for practical sunlight imaging will be discussed separately.

\section*{Acknowledgments}
The authors wish to thank J. Simon and Hui Chen for their helpful discussions. This material is based upon work partially supported by the National Science Foundation and the Maryland Innovation Initiative (MII).
\section*{Author contributions statement}
Y.H. conceived the experiment(s),  J.S. conducted the experiment(s), J.S. and T.P. analyzed the results.  All authors reviewed the manuscript. 
\section*{Additional Information}
\textbf{Competing financial interests} The authors declare no competing financial interests.
\end{document}